\documentstyle[11pt,newpasp,twoside,epsf,epsfig,graphics]{article}
\def\edcomment#1{\iffalse\marginpar{\raggedright\sl#1\/}\else\relax\fi}
\reversemarginpar

\begin{document}
\title{Modulation of Ca II H \& K Emission by Short-Period Planets}
  \author{Evgenya Shkolnik, Gordon A. H. Walker}
\affil{Department of Physics \& Astronomy, University of British Columbia, 6224 Agricultural Road, Vancouver, BC V6T 1Z1, Canada}

\author{David A. Bohlender}
\affil{Herzberg Institute of Astrophysics, 5071 West Saanich Road, Victoria, BC V9E 2E7, Canada}

\begin{abstract}
We have detected modulation of the Ca II H\&K reversal structure in four out of five
51~Peg-type stars whose planets have orbital periods between 3 and 4 days. We observe
two effects in the K-core: (1) a broad 3-\AA~ variation at $\approx$1\% level and (2) changes
on a scale of 0.5 \AA~ ($\sim$1-3\%) in each of the three reversal features.
The nightly variations are coherent in both H and K.  From differential radial velocities
measured to better than 20 m/s, up-to-date phases were extracted.  The enhancements in the reversals tend
to be greatest at the sub-planetary points which may imply that there is a magnetic
interaction between the star's outer layers and the magnetosphere of the planet.  These high-S/N
(500 per pixel in the continuum) and high-resolution (R = 110,000) data are too few to
confirm orbital synchronization. 
\end{abstract}

\section{Introduction}

Current planet detection methods give basic information on the planet: a minimum mass if the orbital inclination is not known, an estimate of surface temperature, and, in the case of the one transiting system, a possible sodium detection in the planet's atmosphere.  There remains a lack of constraints on the planet's structure leaving astronomers to explore new observational probes.

Cuntz, Saar, \& Musielak (2000) suggested that there may be an observable interaction between a parent star and a close-in giant planet.  The effect could be tidal, magnetic, or a combination of the two.  Such an interaction could manifest itself in the form of chromospheric and coronal heating which are predicted to produce a variation of a few percent (Saar \& Cuntz 2001). If the surface
of the star is indeed heated by the planet, then a hot-spot would follow the planet around its orbit, appearing hottest at the sub-planetary point if the interaction is with the planet's magnetosphere, or twice per orbit
if the interaction is tidal. There are several indicators of increased chromospheric activity, one of which is the Ca II H \& K line reversal.  Due to the accessibility of these lines with ground-based telescopes, we chose to monitor the spectral region over several nights for five planetary systems whose orbital periods are between 3.1 and 4.6 days.

\section{Observations}
High-resolution (R = 110,000) and high signal-to-noise (S/N per pixel $\approx$ 500 in the continuum
and 150 in the H \& K cores) spectra were obtained over 3.5 nights at the CFHT with the fiber-fed coud\'{e}
spectrograph. The spectral range spanned 60 \AA~ centered on 3947 \AA. Differential radial velocities were measured to better than 20 m/s allowing current
ephemerides and hence accurate phases ($\pm$0.05) to be determined for each spectrum.

The data showed significant nightly modulations in the Ca II H \& K reversal structure in the spectra of $\tau$ Boo, 51 Peg, HD179949,
HD 209458 and $\upsilon$ And.  For the first four of these stars, the highest enhancement in the K-line emission occurred at
the sub-planetary point where $\phi\sim$~0.  However, for $\upsilon$ And, the modulation was not correlated to orbital phase and was at a smaller level, though
significantly higher than the two standards, $\tau$ Ceti and the Sun.
Table 1 lists the target stars with the RMS of the maximum K-line activity measured relative to each star's normalized mean. As a test of
stability of the spectra, the identical analysis was performed on the aluminum line (3944\AA) which is of comparable depth
to the H \& K absorption features.  The RMS of these are also included in Table 1.


\begin{table}
\begin{center}
\caption{RMS Values of Residuals Relative to Normalized Mean}
\begin{tabular}{lccc}
\\
\tableline
Star		& K-core$^{a}$	& K-line$^{b}$	& Al-line\\
 		& \%	& \%	& \%			\\
\hline
HD 179949	& 0.84	& 0.53	& 0.13			\\
HD 209458	& 1.69	& 1.59	& 0.23			\\
$\tau$ Boo	& 1.14	& 0.97	& 0.25			\\
51 Peg	        & 0.91	& 0.61	& 0.08			\\
$\upsilon$ And	& 0.24	& 0.34	& 0.14			\\
$\tau$ Ceti (Standard)	& 0.03	& 0.17	& 0.08		\\					
\tableline
\tableline

$^{a}$Measured over 0.5 \AA	\\
$^{b}$Measured over 7 \AA	\\

\end{tabular}
\end{center}
\end{table}


The residuals taken from the normalized mean of four nights of HD 209458 and $\tau$~Ceti, a star known not to have a close-in giant planet, are shown in Figure 1. From these, a large 3-\AA~variation with varying intensity is evident along with smaller features (0.5-\AA) superimposed. These smaller features are more evident for HD 179949. An example of the Ca II K-line for HD 179949 is shown in Figure 2 with the residuals. There appear to be changes in the three individual features of the reversal: the central
self-absorption and the blue and red shoulders.  The two spectra taken on different nights but of similar phase ($\phi\sim$0.7) appear to have
a consistent shape while the spectrum at $\phi$=0.07 has a clear enhancement in the center self-reversal.

\begin{figure}[h]
\epsfxsize=5.5in
\epsfig{file=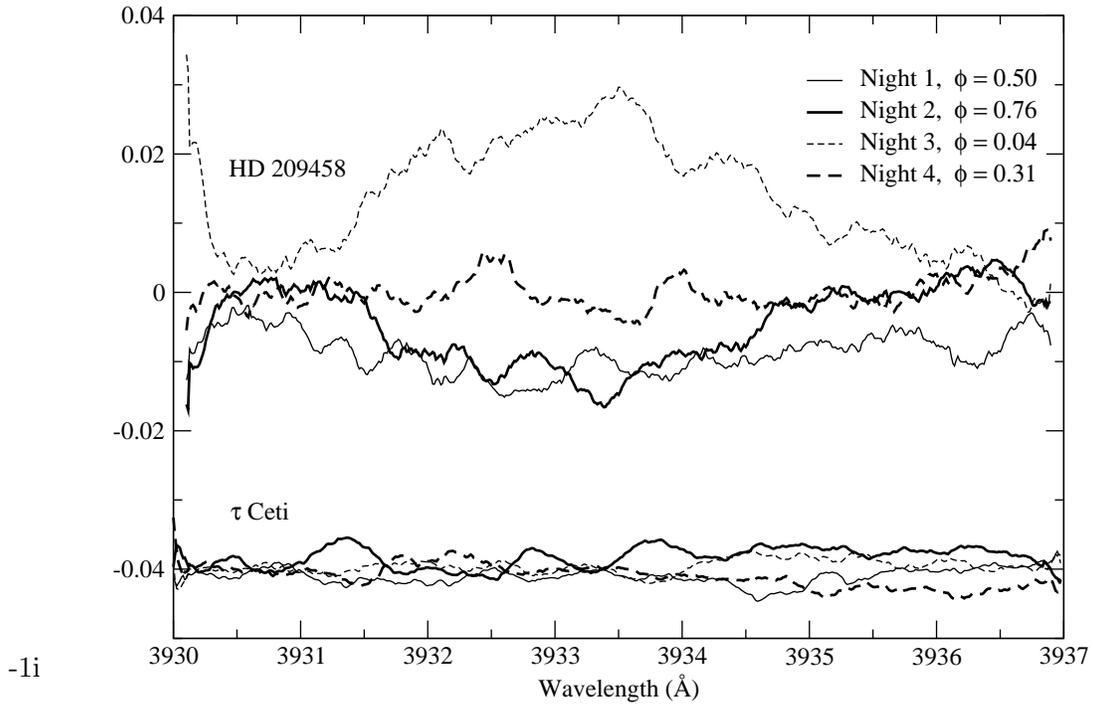,width=3.5in,trim=-1in 0in 0 -0.8in}
\caption{K-line residuals relative to a normalized mean for four nights of HD209458 and  $\tau$~Ceti. Each night is labeled with orbital phase, $\phi$.}
\end{figure}

\begin{figure}[h]
\epsfxsize=5.5in
\epsfig{file=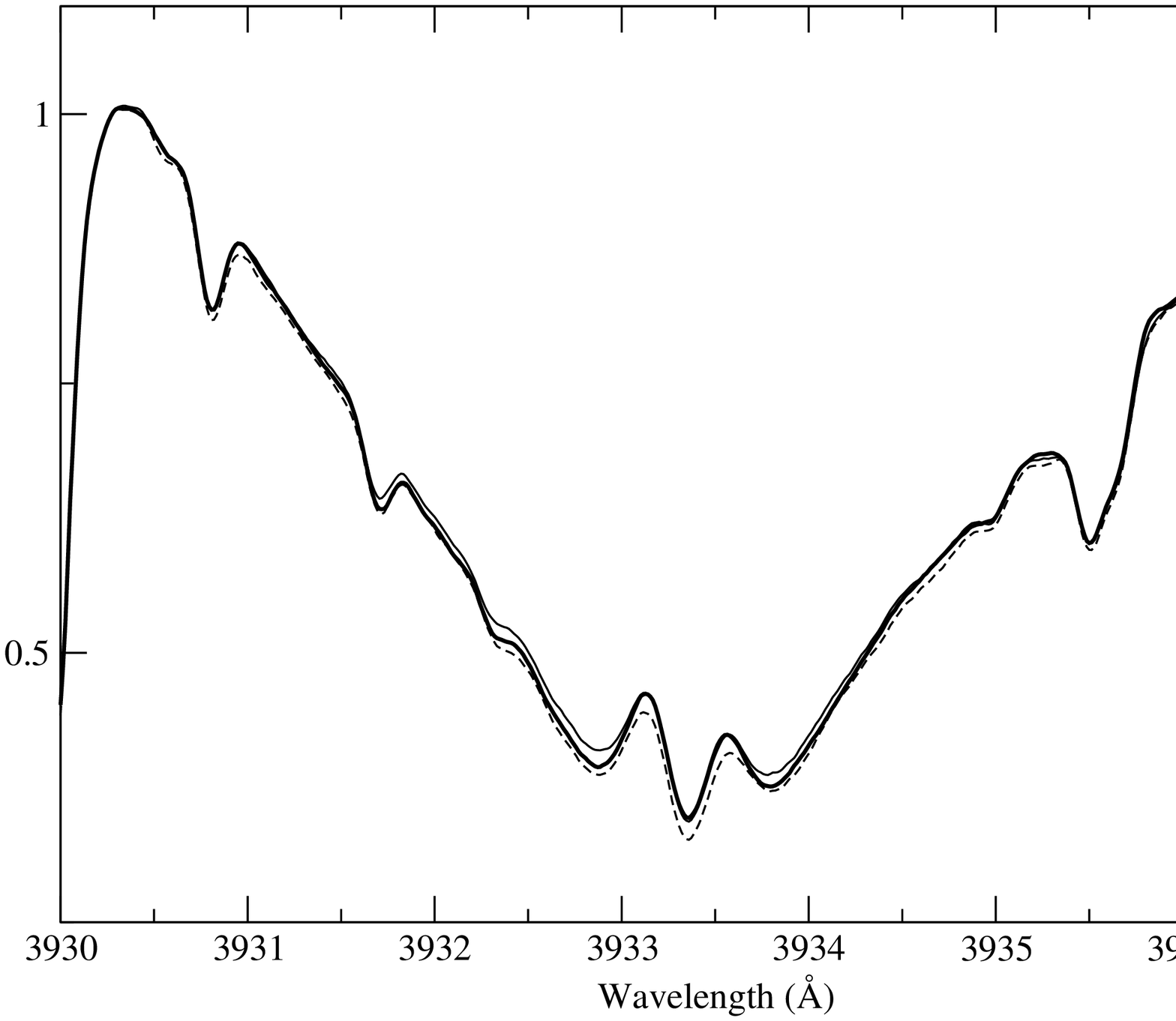,width=3.5in,trim=-1.25in 0in 0 -0.5in}
\epsfxsize=5.5in
\epsfig{file=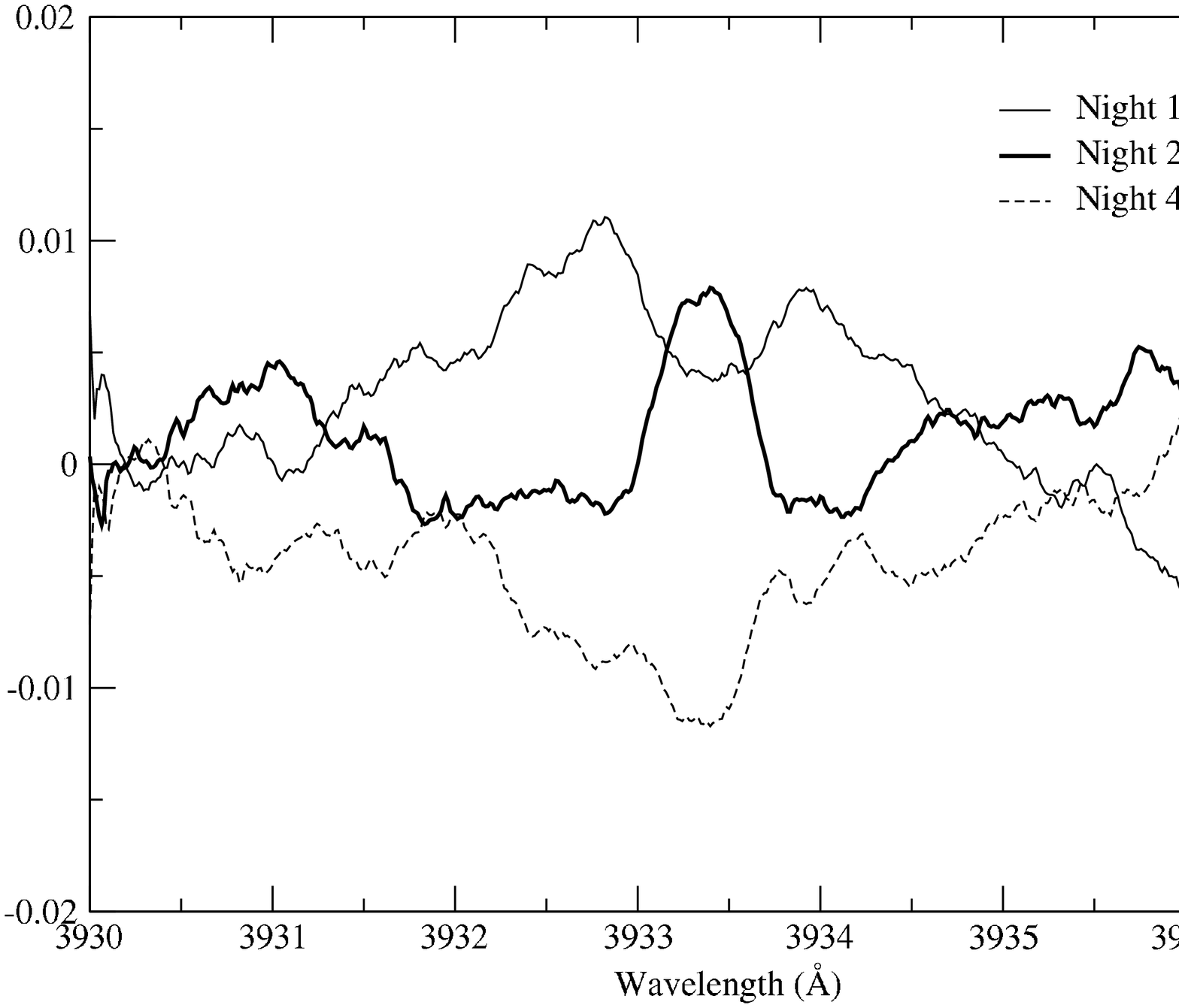,width=3.5in,trim=-1.25in 0in 0 -0.8in}
\caption{The Ca II K-line of HD 179949 for three nights of data (top) with the nightly residuals relative to the normalized mean (bottom).}
\end{figure}

Preliminary evidence indicates that the emission increases once per orbit in four of the five stars near the sub-planetary point, $\phi$=0.
This may suggest that the planet-star interaction is of a magnetic nature and any tidal heating that may be occurring is as yet undetected.
If follow-up observations confirm phase-dependent heating of the stellar chromosphere, then this could be the first evidence
 of an extra-solar planet's magnetosphere.

\end{document}